\documentclass[preprint,showpacs,preprintnumbers,amsmath,amssymb]{revtex4}

\usepackage{graphicx}
\usepackage{dcolumn}
\usepackage{bm}

\begin{document}

\title{Gigantic magnetoelectric effect caused by magnetic-field-induced canted antiferromagnetic-paramagnetic transition in quasi-two-dimensional Ca$_2$CoSi$_2$O$_7$ crystal}

\author{M. Akaki}
\email{m-akaki@sophia.ac.jp}
\affiliation{
Department of Physics, Sophia University, Tokyo 102-8554, Japan
} 

\author{J. Tozawa}%
\affiliation{
Department of Physics, Sophia University, Tokyo 102-8554, Japan
} 

\author{D. Akahoshi}%
\affiliation{
Department of Physics, Sophia University, Tokyo 102-8554, Japan
} 

\author{H. Kuwahara}%
\affiliation{
Department of Physics, Sophia University, Tokyo 102-8554, Japan
} 

\date{\today}

\begin{abstract}
We have investigated the magnetic and dielectric properties of Ca$_2$CoSi$_2$O$_7$ crystal. 
The dielectricity and magnetism of Ca$_2$CoSi$_2$O$_7$ are strongly coupled below a canted antiferromagnetic transition temperature ($T_{\rm N}$): Magnetic fields induce electric polarization below $T_{\rm N}$.
Interestingly, the magnetic-field-induced electric polarization is detected even without poling electric fields.
Below $T_{\rm N}$, a canted antiferromagnetic-paramagnetic transition is induced by magnetic fields. The large magnetocapacitance is observed around $T_{\rm N}$. The origin of the large magnetocapacitance is due to the magnetic-field-induced the canted antiferromagnetic-paramagnetic transition.
\end{abstract}

\maketitle

Since the discovery of the giant magnetoelectric (ME) effect in TbMnO$_3$,\cite{TbMn} multiferroic materials showing simultaneous ferroelectric and magnetic ordering have been attracting much attention, and considerable efforts have been devoted to discover multiferroic materials.
It is generally recognized that spiral (e.g. TbMnO$_3$\cite{Tbspiral}) or up-up-down-down (e.g. HoMnO$_3$\cite{Etype}) spin order is responsible for the giant ME effect. In spiral spin structure, electric polarization is induced due to the inverse Dzyaloshinskii-Moriya (DM) interaction,\cite{Katsura}  while in up-up-down-down spin structure, the origin of electric polarization is explained by exchange-striction mechanism.\cite{Etype, RMn2O5}
The present compound Ca$_2$CoSi$_2$O$_7$ has the same crystal structure as Ba$_2$CoGe$_2$O$_7$ which is a multiferroic material recently reported.\cite{BaCoGe}
As seen from the schematic crystal structure illustrated in Figs.\@ 1 (a) and (b), CoO$_4$ tetrahedra and SiO$_4$ ones are connected with each other through their corners to form two-dimensional layers, and the two-dimensional layers are stacking along the $c$ axis with intervening Ca layers.
The magnetic structure of Ba$_2$CoGe$_2$O$_7$ is different from those of the multiferroic materials recently discovered, i.e., Ba$_2$CoGe$_2$O$_7$ has in-planar antiferromagnetic structure below the N\'eel temperature, $T_{\rm N}$ = 6.7 K,\cite{BaCoGe2, BaCoGe3, BaCoGe4} where ferroelectricity is observed.\cite{BaCoGe} 
The ME effect of Ba$_2$CoGe$_2$O$_7$ is distinguished from those observed in other multiferroic materials: The electric polarization does not flop sharply and it rotates gradually by applying magnetic fields in contrast to the case of TbMnO$_3$.
In this study, we have investigated the ME properties of Ca$_2$CoSi$_2$O$_7$ single crystal, and found out a magnetoelectric effect caused by a magnetic-field-induced canted antiferromagnetic-paramagnetic transition, which is quite different from that of Ba$_2$CoGe$_2$O$_7$ and other multiferroic materials.

The single crystalline sample was grown by the floating zone method. We performed X-ray-diffraction and rocking curve measurements on the obtained crystal at room temperature, and confirmed that the sample has the tetragonal $P\overline{4}2_1 m$\cite{CaCo} structure without any impurity phases or any phase segregation. All specimens used in this study were cut along the crystallographic principal axes into a rectangular shape by means of X-ray back-reflection Laue technique. The magnetization and specific heat were measured using a commercial apparatus (Quantum Design, PPMS). The dielectric constant was measured at 100 kHz using an {\it LCR} meter (Agilent, 4284A). The spontaneous electric polarization was obtained by the accumulation of a pyroelectric current for temperature scans and a displacement current for magnetic-field ones.

We present in Fig.\@ 1 the temperature dependence of (c) the dielectric constant and (d) magnetization for Ca$_2$CoSi$_2$O$_7$ crystal. Ca$_2$CoSi$_2$O$_7$ undergoes a canted antiferromagnetic transition at 5.7 K, where the magnetization parallel to the $a$ axis ($M_a$) abruptly increases, just like the case of Ba$_2$CoGe$_2$O$_7$.\cite{BaCoGe}
This magnetic transition does not accompany any thermal hysteresis, meaning that it is of second-order.
We confirmed that Ca$_2$CoSi$_2$O$_7$ has a collinear spin structure similar to that of Ba$_2$CoGe$_2$O$_7$ by preliminary neutron diffraction experiments.\cite{Kubota}
All spins lie within the $ab$ plane, and the nearest spins along the [110] direction are ordered antiferromagnetically.
Therefore, the magnetic moment perpendicular to the $c$ axis is probably due to spin canting caused by DM interaction. 
In contrast, the magnetization parallel to the $c$ axis ($M_c$) exhibits no anomaly around a canted antiferromagnetic transition temperature ($T_{\rm N}$) of 5.7 K. 
The dielectric constant parallel to the $c$ axis ($\varepsilon_c$) shows a slight jump at $T_{\rm N}$, while the dielectric constant parallel to the $a$ axis ($\varepsilon_a$) does not show any anomaly. These results suggest that the magnetism and dielectricity are substantially coupled in Ca$_2$CoSi$_2$O$_7$.

Figure 2 displays the temperature dependence of (a) the dielectric constant, (b) spontaneous electric polarization, and (c) specific heat for Ca$_2$CoSi$_2$O$_7$ crystal in magnetic fields. In the absence of magnetic fields, $\varepsilon_a$ does not show any anomaly (Fig.\@ 1 (c)), nor is electric polarization observed. 
When applying magnetic fields parallel to the $c$ axis ($H\parallel c$), a peak emerges  around $T_{\rm N}$ in the temperature dependence of $\varepsilon_a$. With an increase in applied magnetic fields, the peak position of $\varepsilon_a$ shifts to lower temperatures, but the peak value is gradually increased (Fig.\@ 2 (a)). 
Specific heat measurements show that applying magnetic fields reduces $T_{\rm N}$ (Fig.\@ 2 (c)).
Accompanying the emergence of the $\varepsilon_a$-peak, the electric polarization is observed below $T_{\rm N}$ along the $a$ axis, but not along the $c$ axis\cite{epsironc}.
The electric polarization is developing with increasing applied magnetic fields well below $T_{\rm N}$. 
However, near $T_{\rm N}$ it is suppressed in $H>4$ T (Fig.\@ 2 (b)). 
This is simply because $T_{\rm N}$, that is, the onset temperature of the electric polarization shifts to lower temperatures by applying magnetic fields.
When applying magnetic fields parallel to the $a$ axis, electric polarization emerges along the $c$ axis (not shown), that is to say, the direction of the electric polarization is always perpendicular to that of applied magnetic fields.

Figure 3 (a) displays the isothermal magnetocapacitance ($\Delta \varepsilon (H)/\varepsilon (0) \equiv [\varepsilon (H)-\varepsilon (0)]/\varepsilon (0)$) curves at several fixed temperatures around $T_{\rm N}$. At 5.9 K, immediately above $T_{\rm N}$, the magnetocapacitance increases with an increase in magnetic fields, and it reaches about 2 \% around $H=4$ T.
Below $T_{\rm N}$, the magnetocapacitance shows a somewhat different behavior. At 5.7 K, the magnetocapacitance is almost negligible at lower magnetic fields. Around $H=2$ T, the magnetocapacitance shows a sudden increase and takes a maximum value around $H=4$ T.
As temperature is lowered, the maximum value of the magnetocapacitance is considerably enhanced: $\Delta \varepsilon (H)/\varepsilon (0)$ reaches 13 \% at 5.1 K at $H=8$ T. The magnetocapacitance observed in Ca$_2$CoSi$_2$O$_7$ is relatively large compared with those of other multiferroic materials recently discovered (e.g., TbMnO$_3$: 10 \%\cite{TbMn}, MnWO$_4$: 4 \%\cite{MnW}, LiCu$_2$O$_2$: 0.4 \%\cite{LiCu}). 
The large magnetocapacitance is unambiguous evidence of strong coupling between the dielectricity and magnetism in Ca$_2$CoSi$_2$O$_7$.
Figure 3 (b) shows the magnetic field dependence of the electric polarization. At 5.7 K, the electric polarization monotonically increases with increasing magnetic fields up to $H\approx 3$ T, where it takes a maximum value. 
With decreasing temperature, the peak position shifts to higher magnetic fields, and the maximum value of the electric polarization is further evolved.
Note that the peak position almost coincides with the point where the magnetocapacitance shows an abrupt increase. 
As seen from Fig.\@ 3 (c), the magnetization curve below $T_{\rm N}$ almost linearly depends on magnetic fields, and does not show any clear inflection point, suggesting that neither spin flop nor first-order phase transition is induced by magnetic fields. The field derivative of the magnetization ($dM/dH$) slightly changes around $H=4$ T (inset of Fig.\@ 3 (c)), where canted antiferromagnetic state turns into paramagnetic one.

As seen from the above results, although Ca$_2$CoSi$_2$O$_7$ has the similar crystal and magnetic structures to Ba$_2$CoGe$_2$O$_7$, the ME properties are rather different from those of Ba$_2$CoGe$_2$O$_7$. The significant differences are as follows. First, the electric polarization of Ca$_2$CoSi$_2$O$_7$ is never observed unless magnetic field is applied.
In Ba$_2$CoGe$_2$O$_7$, electric polarization is obviously observed even in $H=0$. 
Second, the direction of the electric polarization of Ca$_2$CoSi$_2$O$_7$ is always perpendicular to that of applied magnetic fields.
In contrast, the electric polarization of Ba$_2$CoGe$_2$O$_7$ along the $c$ axis is enhanced by magnetic fields parallel to the same axis.
The large magnetocapacitance is found in Ca$_2$CoSi$_2$O$_7$ as well as Ba$_2$CoGe$_2$O$_7$. However, the origin is quite different from each other.
In the case of Ba$_2$CoGe$_2$O$_7$,the dielectric constant largely varies when the direction of the electric polarization rotates by applying magnetic fields, which explains the origin of the large magnetocapacitance of this compound.
In Ca$_2$CoSi$_2$O$_7$, the $\varepsilon_a$-peak temperature associated with $T_{\rm N}$ is reduced across a given temperature with increasing magnetic fields, resulting in the large isothermal magnetocapacitance at that temperature.
That is, the large magnetocapacitance of Ca$_2$CoSi$_2$O$_7$ is caused by the magnetic-field-induced canted antiferromagnetic-paramagnetic transition accompanying the large change in $\varepsilon_a$.
Another striking characteristic of Ca$_2$CoSi$_2$O$_7$ is that the magnetic-field-induced electric polarization is observed even without poling electric fields, which cannot be reversed by sign change of DC electric fields.
The electric polarization is linearly induced with increasing magnetic fields below 5.7 K (Fig.\@ 3 (b)). 
Since Ca$_2$CoSi$_2$O$_7$ has a commensurate and collinear magnetic order,\cite{Kubota}  the emergence of the electric polarization cannot be explained by spin-current model.\cite{Katsura}
Ca$_2$CoSi$_2$O$_7$ has the space group, $P\overline{4}2_1 m$ at room temperature,\cite{CaCo} which means that Ca$_2$CoSi$_2$O$_7$ is piezoelectric. The electric polarization does not appear below $T_{\rm N}$ in a zero magnetic field. Therefore, it is reasonable to consider that Ca$_2$CoSi$_2$O$_7$ remains piezoelectric below $T_{\rm N}$.
Lattice strain through magnetostriction likely causes the electric polarization of Ca$_2$CoSi$_2$O$_7$.

In summary, we have investigated the ME properties of Ca$_2$CoSi$_2$O$_7$ crystal. 
The ME behavior is distinguished from that of other multiferroics.
The canted antiferromagnetic transition occurs at 5.7 K, below which the dielectricity and magnetism are strongly coupled.
The large magnetocapacitance is observed below $T_{\rm N}$, which is caused by the magnetic-field-induced canted antiferromagnetic-paramagnetic transition accompanying the large change in $\varepsilon_a$.
Below $T_{\rm N}$, the electric polarization is induced perpendicular to the direction of magnetic fields. 
Interestingly, the electric polarization can be observed even without poling electric fields.
The present results provide significant information on technical application of multiferroics.

This work was supported by Grant-in-Aid for JSPS Fellows and Scientific Research (C) from Japan Society for Promotion of Science.

\newpage

\begin{figure}[tb]
\begin{center}
\includegraphics[width=0.45\textwidth,clip]{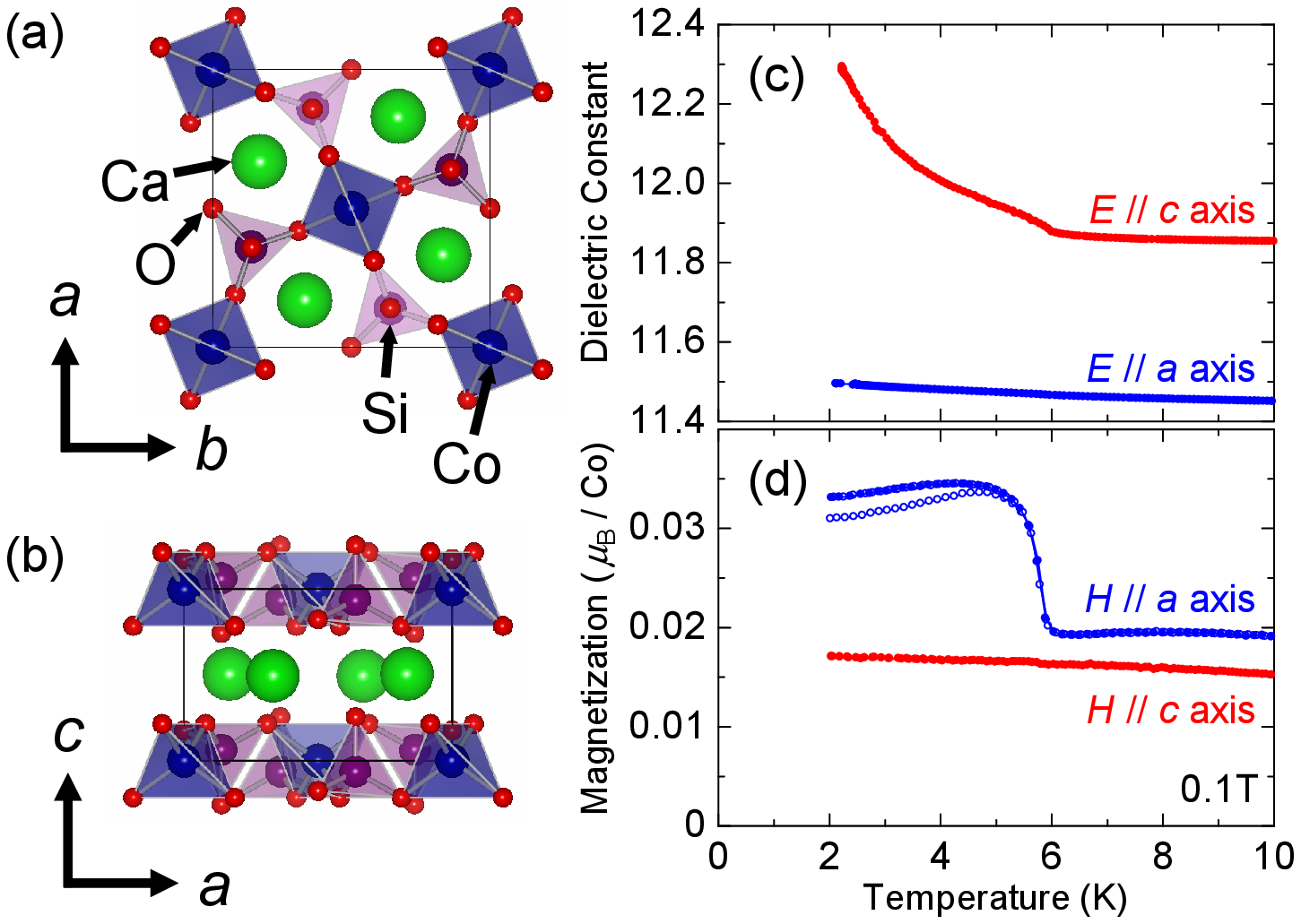}
\end{center}
\caption{(Color online) Schematic crystal structure of  Ca$_2$CoSi$_2$O$_7$ projected onto the (a) $ab$ plane and (b) $ac$ plane. Temperature dependence of (c) dielectric constant and (d) magnetization in Ca$_2$CoSi$_2$O$_7$. The magnetization was measured in warming scan after zero-field-cooling (ZFC: open symbols) and field-cooling (FC: closed symbols). A magnetic field of 0.1 T was applied parallel to the $a$ and $c$ axes.}
\label{f1}
\end{figure}

\begin{figure}[tb]
\begin{center}
\includegraphics[width=0.38\textwidth,clip]{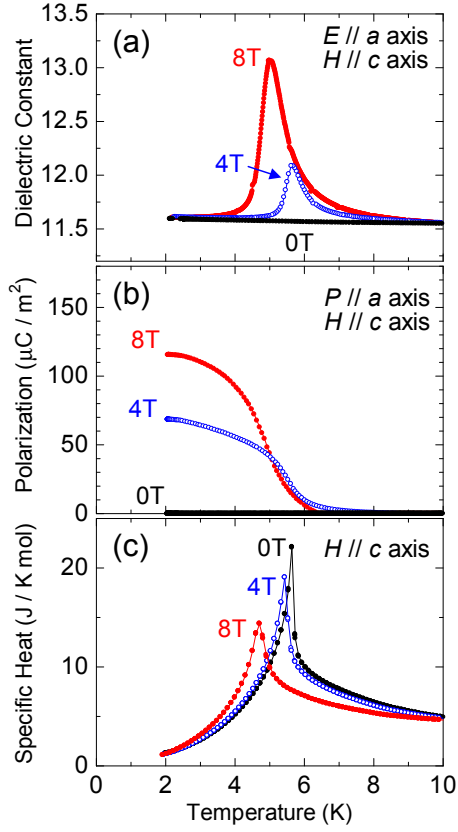}
\end{center}
\caption{(Color online) Temperature dependence of (a) dielectric constant, (b) spontaneous electric polarization, and (c) specific heat of Ca$_2$CoSi$_2$O$_7$ in magnetic fields parallel to the $c$ axis. The dielectric constant and spontaneous electric polarization measured along the $a$ axis.}
\label{f2}
\end{figure}

\begin{figure}[tb]
\begin{center}
\includegraphics[width=0.38\textwidth,clip]{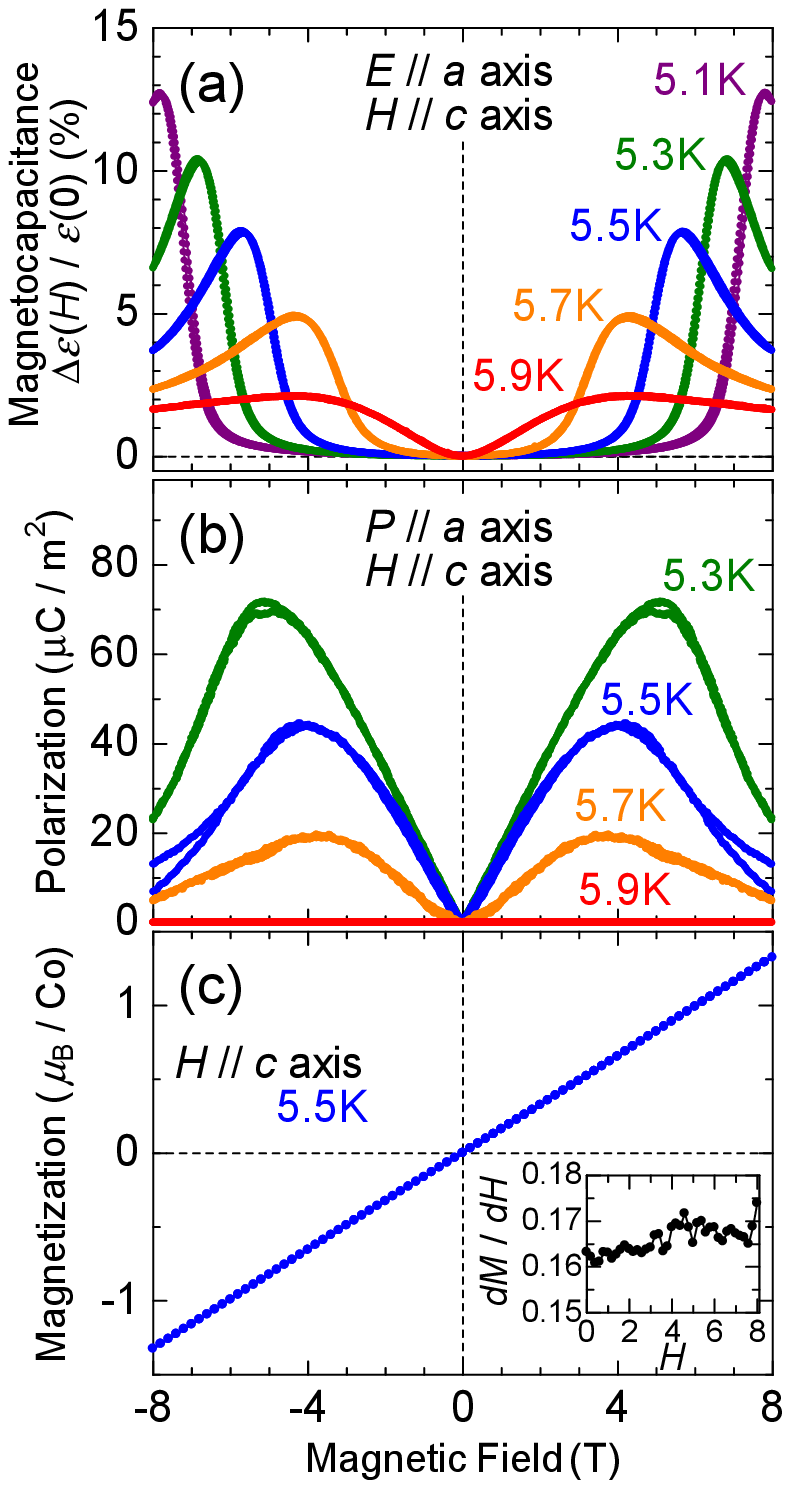}
\end{center}
\caption{(Color online) (a) normalized magnetocapacitance by zero field, (b) spontaneous electric polarization, and (c) magnetization of Ca$_2$CoSi$_2$O$_7$ crystals as a function of an external magnetic field parallel to the $c$ axis at several fixed temperatures. The inset shows the field derivative of the magnetization.}
\label{f3}
\end{figure}


\begin{thebibliography}{99}
\bibitem{epsironc} In a zero magnetic field, $\varepsilon_c$ shows a slight jump at $T_{\rm N}$ (Fig.\@ 1(c)). Application of $H\parallel c$ suppresses the $\varepsilon_c$-jump, and no anomaly is seen around  $T_{\rm N}$ under higher magnetic fields. Electric polarization cannot be observed along the $c$ axis by applying $H\parallel c$.
\end{thebibliography}
\end{document}